%% file: proceedingsf.tex
\documentclass[11pt]{article}
\usepackage{epsfig}
\usepackage{amsmath,amssymb}
\usepackage{graphicx} 
\input econfmacros.tex
\def\Title#1{ {\Large {\bf #1} }}

\begin{document}

\Title{A Review on the Scalar Field/ Bose-Einstein Condensate Dark Matter Model}

\bigskip\bigskip


\begin{raggedright}  

{{\bf Abril Su\'arez\index{Su\'arez A.}}\\
Departamento de F\'isica, Centro de Investigaciones y de Estudios Avanzados del IPN, A.P.\\
14-740, 07000 M\'exico D.F., M\'exico\\
E-mail: asuarez@fis.cinvestav.mx}\\
\bigskip\bigskip
{{\bf Victor H. Robles\index{Robles V.H.}}\\
Departamento de F\'isica, Centro de Investigaciones y de Estudios Avanzados del IPN, A.P.\\
14-740, 07000 M\'exico D.F., M\'exico\\
E-mail: vrobles@fis.cinvestav.mx}\\
\bigskip\bigskip

{{\bf Tonatiuh Matos\index{Matos T.}}\\
Departamento de F\'isica, Centro de Investigaciones y de Estudios Avanzados del IPN, A.P.\\
14-740, 07000 M\'exico D.F., M\'exico\\
E-mail: tmatos@fis.cinvestav.mx}\\
\bigskip\bigskip
\end{raggedright}

\abstract{We review the work done so far aimed at modeling in an alternative way the dark matter in the Universe: the  scalar field/ Bose-Einstein condensate dark matter (SFDM/BEC) model. We discuss a number of important achievements and characteristics of the model. We also describe some of our most recent results and predictions of the model compared to those of the standard model of $\Lambda$CDM.}

\section{Introduction}

It is a pleasure for us to review the different theoretical basis of the SFDM/BEC model as a dark matter (DM) candidate. 
We think this review is important for mainly two reasons: the considerable progress in the model since it was proposed as a 
serious candidate to the dark matter paradigm, improved theoretical understanding of the nature of the DM  and the significant 
advances in the cosmological and astronomical observations are leading us to put more constraints and will allow us to test the model and decide 
if it can still stand as a viable DM paradigm or if it should be discarded.

Recent observations of the Universe have found that only 4 percent of the total content of the Universe is baryonic matter, 
being 22 percent of the rest remaining non-baryonic dark matter (DM) and the rest in some form of cosmological constant.

The incorporation of a new kind of DM different from that proposed by the standard model, also known by $\Lambda$-Cold Dark Matter model ($\Lambda$CDM), 
into the Big Bang Theory holds out the possibility of giving alternative answers to some of the unsolved issues of the standard 
cosmological model. Several authors have proposed interesting alternatives in where they try to solve the difficulties 
that the $\Lambda$CDM scenario seems not to solve. In fact, the alternative Scalar Field Dark Matter (SFDM/BEC) scenario has 
received much attention in the last few years. The main idea is simple, in this models the nature of the DM is completely determined by a 
fundamental scalar field $\Phi$ \cite{b1}.

The idea was first considered in \cite{b3} and independently introduced in \cite{b1,b4,b5} suggesting bosonic dark matter as a 
model for galactic halos, see also \cite{b6}. In the SFDM/BEC model, DM halos can be described, in the non-relativistic regime, 
as Newtonian gravitational condensates made up of ultra-light bosons that condense into a single macroscopic wave function.

Several authors have introduced a dynamical scalar field with a certain potential $V(\Phi)$ as a candidate to be the dark matter, although  there is not yet an agreement for the correct form of the potential field. Other interesting works consider a single scalar field to unify the description of dark matter, dark energy and inflation \cite{b7,b8,b9}. 

Different issues of the cosmological behavior of the SFDM/BEC model have been studied in a wide variety of approaches, 
see for example \cite{b10,b11,b12,b13,b14,b15,b16,b17,b18,b19,b20,b21,b22,b23,b24,b25,b26}. For example, \cite{b12} proposed fuzzy 
dark matter composed of ultra-light scalar particles initially in the form of a BEC. Recently \cite{b20,b21} developed a further 
analysis of the cosmological dynamics of SFDM/BEC as well as the evolution of their fluctuations (see also \cite{b25}). In the same 
direction, \cite{b19,b27} studied the growth of scalar fluctuations and the formations of large-scale structure within a fluid and a 
field approach for the SFDM/BEC model.

In addition, many numerical simulations have been performed to study the gravitational collapse of the SFDM/BEC model \cite{b28,b29,b30,b31,b32,b33,b34,b35,b36,b37}.  \cite{b38,b39} found an approximate analytical expression and numerical solutions of the mass-radius relation of SFDM/BEC halos. Recently, \cite{b40} gave constraints on the boson mass to form and maintain more than one vortex in SFDM/BEC halos. These constraints are in agreement with the ultra-light mass found in previous works (see for example, \cite{b41}). Lately, \cite{b42} performed N-body simulations to study the dynamics observed in the Ursa Minor dwarf galaxy. They modeled the dark matter halo of Ursa Minor as a SFDM/BEC halo to establish constraints for the bosons mass. Moreover, they introduced a dynamical friction analysis within the SFDM/BEC model to study the wide distribution of globular clusters in Fornax. An overall good agreement is found for the ultra-light mass of bosonic dark matter.

In this paper we review a number or results and important characteristics of the SFDM/BEC model, its dynamical mechanism and some of 
its predictions. We also discuss some of the trending topics nowadays in the subject which attempt to predict and ask how well the 
model is achieving its goals. As we will see, a number of them studies results which are in reasonable agreement with the general features 
required by the theory and the data.

The outline of the paper is as follows. In Section 1 we have given a brief synopsis on the current state-of-the-art of the model. 
In the Section 2 we describe the dark matter paradigm and briefly resume the standard model of cosmology. 
In Section 3 and 4 we describe in some detail why Scalar Field/Bose-Einstein condensate Dark Matter (SFDM/BEC) can be a good 
alternative candidate to be the dark matter in our Universe, we summarize some representative papers  for this sections, and 
in Section 5 we include topics for future works and our conclusions.

\section{The Dark Matter Paradigm and the Standard Model of Cosmology}

In this millennium, new technologies are opening wider windows to explore our Universe. For some time we could only relay on 
inaccurate evidence found in the local neighborhood of our galaxy to infer the history of our Universe, now it turns it is possible 
for us to see the evolution of the Universe as far as 100,000 years after the Big Bang and in more detail.

With these advances, nowadays some inquires of our cosmic evolution can be determined by giving an answer to question like: How much matter is in the Universe?

Since the discovery of the expansion of the Universe done by Hubble and Slipher \cite{b59} in the 1920's, the common believe had 
been that all energy in the Universe was in the form of radiation and ordinary matter (electrons, protons, neutrons, etc.). 
Over the past few decades, theories concerning the stability of galaxies (\cite{b57, b60}) indicated that most of the mass in our 
Universe is dark (i.e., it does not emits or absorbs light \cite{b61,b62}), therefore resulting unobservable by telescopes. The 
suggestion that "dark matter" may form a large fraction of the density in the Universe was raised by Zwicky in 1937. Back then he 
used the virial theorem to obtain the average mass of galaxies within the Coma cluster and obtained a value much larger than the 
mass of the luminous material, he then realized that some mass was "missing" in order to account for observations. This missing mass 
problem was confirmed many years later by more accurate measurements of rotation curves of disc galaxies,  \cite{b49, b48, b57, b58}. 
The rotation curves of neutral hydrogen clouds in spiral galaxies measured by the Doppler effect are now found to be roughly flat with 
a typical rotation velocity equal to $v_{\infty}\sim 200$km/s up to the maximum observed radius of about 50kpc. With these observations 
the mass profile results much more extended than the common distributions which typically converge within $\sim 10$kpc. This  would imply that galaxies might be surrounded by an extended halo of dark matter whose mass $M(r)\sim rv_{\infty}^2/G$ increases linearly with radius (here $r$ is the radius and $G$ Newton's gravitational constant).

In the 1980's, the proposal of dark matter found its basis in the so called "inflationary scenario" \cite{b63,b64,b65}, a theory of the 
first $10^{-30}$ seconds developed to give answer to several questions left unanswered by the Big Bang model, like for example: Why is 
the Universe so homogeneous and isotropic? and; Where did the initial homogeneities that gave rise to the structures we see today came 
from? \cite{b66,b67,b68,b69,b70}. The inflationary theory predicts that the Universe is spatially flat; which according to Einstein's 
theory of general relativity, this fixes the total energy density of the Universe making it equal to the critical value, $\rho_c\equiv 3H_0^2/8\pi G\sim 1.7\times 10^{-29}\mbox{g cm}^3$, where $H_0$ is the current value of the Hubble parameter. 

Several astrophysical observations of distant type Ia supernovae have also shown that the content of the Universe is made of about 
74\% of dark energy, 22\% of dark matter and 4\% of baryonic (visible) matter, \cite{b71,b72,b73,b53,b54,b55}. Inflation thus seemed to call for dark matter.

It then results that most matter in the Universe is non-luminous. The observed flatness of the galactic rotation curved indicating the 
presence of dark matter halos around galaxies \cite{b57}. Summed to these evidences are the observations of the cosmic microwave 
background (CMB) anisotropies (\cite{b88,b85}) combined with large-scale structure and type Ia supernova luminosity data 
(\cite{b53,b89, b90,b54}) which all together constrain the cosmological parameters also finding once more that visible matter contributes only about 4\% of the energy density of the Universe, gravitational lensing \cite{b86} and X-ray spectra (\cite{b87,b81,b62, b82,b83}) in elliptical galaxies, and the high velocity dispersion and gas temperature in clusters of galaxies (\cite{b62,b84}), all of them leading to a picture in which galaxies are composed of a luminous galactic disk surrounded by a galactic halo of dark matter.  Also the relative contribution of the dark matter component is usually specified in terms of the mass-to-light ratio, M/L; which reflects the total amount of mass relative to the total light within a given scale. The increase on this ratio suggests that there is relatively more dark than luminous matter with increasing scale \cite{b57}. This has led to the general belief that clusters have more dark matter per unit luminosity than individual galaxies and that superclusters may have even more. This widely accepted monotonic increase of M/L with scale determines to a large extent the prevalent views about the location of the dark matter and the total mass density of the Universe. Recent studies of the dependence of the mass-to-light ratio on scale indicate that M/L is nearly constant on large scales ranging up to supercluster size (10Mpc), suggesting no additional dark matter is tucked away on large scales \cite{b76}.  More recently, a clear separation between the center of baryonic matter and the total center of mass was observed in the Bullet cluster (\cite{b43}) and later in other galaxy cluster collisions (\cite{b91}). 

The observational evidence for dark matter continues to grow, and particle physicists have proposed various particles, motivated by supersymmetry and unified theories, that could reasonably explain it.  These observations reinforce the claim that dark matter is indeed composed of weakly interacting particles and is not a modification of gravity. However, even taking into account all these results the properties of dark matter are still mysterious.

It then also results that an important question in cosmology has to do with knowing the nature of the so far undefined one quarter part of the content of the Universe, the dark matter. As mentioned before, the cosmological observations seem to support the idea that dark matter can be made of some kind of non-baryonic, non-relativistic and weakly interactive massive particle. Many efforts trying to give an answer to this question have been done in the past few decades, mainly motivated by the idea that the answer will probably change our understanding of the Universe and its dynamics. One of the explanations for DM is the SFDM model.

In the Standard Model of cosmology, the total energy density of the Universe is dominated today by the densities of two components: the "dark matter" which has an attractive gravitational effect like usual matter and the "dark energy" which can be considered as a kind of vacuum energy with a negative pressure, which seems constant today (i.e., a cosmological constant, $\Lambda$). Although the real nature of these two components remains unknown, in the standard model dark matter is generally modeled as a system of collisionless particles. This is known as the "$\Lambda$ Cold Dark Matter" model, which predicts that the Universe contains primarily cold neutral weakly interactive massive particles (WIMPs) which are non-baryonic (\cite{b74, b75}), pressureless, behave like a cold gas, one beyond those existing in the Standard Model of particle physics and have not yet been detected in accelerators or specialized indirect searches,  in particular, the lightest supersymmetric particles, the most popular of which is the neutralino, with a particle mass of the order of 100GeV. Efforts are underway to measure the presence of these particles, but no direct detection has yet been reported.

In order to explain observational data, the $\Lambda$CDM model was developed, \cite{b51,b52} and it is the most simple possibility. In the Standard Model of cosmology the matter component $\Omega_M\sim 26$\% of the Universe decomposes itself into baryons, neutrinos, etc., and cold dark matter which would be responsible for the formation of structure in the Universe. Observations indicate that stars and dust (baryons) represent something like 0.4\% of the total content of matter in the Universe. The measurements of neutrino masses indicate that these contribute nearly with the same amount as matter. In other words, $\Omega_M=\Omega_m+\Omega_{DM}=\Omega_b+\Omega_{\nu}+...+\Omega_{DM}\sim\Omega_m+\Omega_{CDM}$, where $\Omega_{CDM}$ represents the cold dark matter part of the matter contributions, and has a value of $\Omega_{CDM}\sim 0.22$. The value of the amount of baryonic matter is in accordance with the limits imposed by nucleosynthesis, \cite{b2}. This model then considers a flat Universe ($\Omega_{\Lambda}+\Omega_M\equiv 1$) with $96$\% of unknown matter but which is of great importance in the cosmological context. It also supposes a homogeneous and isotropic Universe which evolution can be best described today by Friedmann's equations coming from general relativity and whose main ingredients can be described by fluids with characteristics very similar to those we see in our Universe. We now know that the Universe is not exactly homogeneous and isotropic, but the standard model does give a framework within which the evolution of structures such as galaxies or clusters of galaxies can be studied with their origins coming from small fluctuations in the density of the early Universe. The model assumes a "scale-invariant" spectrum of initial density fluctuations, a spectrum in which the magnitude of the inhomogeneity is the same on all length scales, again as predicted by standard inflationary cosmology \cite{b67,b68,b69,b71}.  Moreover, $\Lambda$CDM seems to be until today the most successful model fitting current cosmological observations \cite{b47}.

The $\Lambda$CDM model successfully describes the accelerated expansion of the Universe, it explains the Cosmic Microwave Background 
radiation in great detail and provides a framework within which one can understand the large-scale isotropy of the Universe, it also 
describes the important characteristics of the origin, nature and evolution of the density fluctuations which are believed to give 
rise to galaxies and other cosmic structures, the Lyman-$\alpha$ forest, the large scale matter distribution, and the main aspects of 
the formation an the evolution of virialized cosmological objects. So far the $\Lambda$CDM model is consistent with the observed 
cluster abundance at $z\sim 0$, it then predicts a relatively little change in the number density of rich clusters as a function of redshift because, due to the low matter density, hardly any structure growth has occurred since $z\sim1$ . The $\Lambda$CDM model can then be "forced" to agree approximately with both the cluster abundance on small scales and the CMB fluctuations on large scales by tilting the power spectrum (by about 30\%) from its standard shape. This tilted variant of the $\Lambda$CDM model then results nearly consistent with observations. The power spectra of $\Lambda$CDM can then be normalized so that it agrees with both the CMB and cluster observations. But as the estimates of the cold dark matter density become more precise, it becomes even more imperative for its composition to be identified.

There remain, however, certain conflicts at galactic scales, like the cusp profile of central densities in galactic halos, the overpopulation of substructures predicted by N-body numerical simulations which are an order of magnitude larger than what has been observed, among others, see for example \cite{b43, b44,b45,b46}. And until today the nature of the dark matter that binds galaxies remains an open question.   

\section{Why Scalar Field Dark Matter?}

In the big bang model, gravity plays an essential role: it collects the dark matter in concentrated regions called 'dark matter halos'. 
Within these large dark matter halos, the baryons  are believed to be so dense that they radiate enough energy to collapse into galaxies
 and stars. The most massive halos, hosts for the brightest galaxies, are formed in regions with the highest local mass density. Less massive halos, hosts for the less bright galaxies, appear in regions with low local densities \cite{b92}. 
These situations appear to be the same as in our extragalactic neighborhood, but there are still problems. Despite all its successful 
achievements the $\Lambda$CDM model requires further considerations. 

The $\Lambda$CDM paradigm faces several challenges to explain observations at galactic scales, such as the central densities of dark 
halos, dwarfs and Low Surface Brightness (LSB) galaxies, the excess of satellite galaxies predicted by N-body simulations,
the formation of bars in disc galaxies, etc.\cite{b45,b43,b46}. In other words, there is not a match between $\Lambda$CDM predictions at galactic scales and what is being observed. Problems with an otherwise successful model are often the key to a new and deeper understanding.

Observations point out to a better understanding of the theory beginning with the Local Void, which contains just a few galaxies that are larger than expected. This problem would be solved if structure grew faster than in the standard theory, therefore filling the local void and giving rise to more matter in the surroundings (\cite{b92}).

Another problem arises for the so called pure disk galaxies, which do not appear in numerical simulations of structure formation in the Standard Model. These problems would be solved again if the structure grew faster than it does in the standard paradigm \cite{b92}. 

On the other hand, \cite{b93} also found that the collision velocity of 3000km/s at R$_{200}$ for the Bullet Cluster is very unlikely 
within the $\Lambda$CDM paradigm, which moves it to a challenge for the Standard Model of cosmology. 

A final example of inconsistencies can be seen in a paper by \cite{b94}, who found anomalies in the mass power spectrum obtained by the SDSS and the one obtained with the $\Lambda$CDM model, i.e., anomalies in the predicted large-scale structure of the Universe.With these and other results it seems necessary to change the $\Lambda$CDM paradigm to try and explain the formation of structure in the Universe.

Given these discrepancies, it seems necessary to explore alternatives to the paradigm of structure formation.  These are some of the reasons why we need to look for alternative candidates that can explain the structure formation at cosmological level, the observed amount of dwarf galaxies, and the dark matter density profiles in the core of galaxies. Recently, several alternative models have been proposed.

One of them invokes a scalar field as dark matter in the Universe \cite{b22,b24}. This model supposes that dark matter is a real or complex scalar field $\Phi$ minimally coupled to gravity, endowed with a scalar potential $V(\Phi)$ and that at some temperature it only interacts gravitationally with the rest of the matter. This scalar field can be added to the particles standard model lagrangian or to the general relativity one, supposing that the coupling constant with the rest of the matter is very small. It has been also suggested that this scalar field can be derived from higher dimensional theories. 
It has also been proposed that this dark matter scalar field, i.e., this spin-0 fundamental interaction, could lead to the formation of Bose-Einstein condensate in the way of cosmic structure (\cite{b10,b6,b20,b21,b12}) with an ultra-light mass of order $m\sim 10^{-22}$eV. From this mass it follows that the critical temperature of condensation $T_c\sim 1/m^{5/3\sim}$TeV is very high, therefore, they may form Bose-Einstein condensate drops very early in the Universe \cite{b22} that behave as cold DM. Lee and Koh \cite{b5}, and independently Matos and Guzm\'an \cite{b1}, suggested bosonic dark matter as a model for galactic halos. In addition, the Compton length $\lambda_c=2\pi\hbar/m$ associated to this boson results of about $\sim$kpc, and corresponds to the dark halo size of typical galaxies in the Universe. Thus, it has been sopported that these drops are the halos of galaxies (see \cite{b10}), i.e., that halos are huge drops of SF. In a recent paper, Ure\~{n}a \cite{b14} studied the conditions for the formation of a SFDM/BEC in the Universe, also concluding that SFDM/BEC particles must be ultra-light bosons.

In the SFDM model the initial halos of galaxies do not form hierarchically, they are formed at the same time and in the same way when the Universe reaches the critical temperature of condensation of the SF. From this it follows that galaxies can share some properties because they formed in the same manner and at the same moment \cite{b95}. 
Therefore, from this paradigm we have to expect that there exists well formed galaxy halos at higher redshifts than in the $\Lambda$CDM model. Recently Su\'arez and Matos \cite{b19} developed a hydrodynamical approach for the structure formation in the Universe with the scalar potential $V(\Phi)=m^2\Phi^2/2+\lambda\Phi^4/4$. They found that when $\lambda=0$ the evolution of perturbations of the SFDM model compared to those of $\Lambda$CDM are identical. They also showed that this potential can lead to the early formation of gravitational structures in the Universe depending on the sign of the self-interaction parameter $\lambda$.
 
The most simple model having both an exponential behavior and a minimum is a $cosh$-like potential.  Another interesting work was done by \cite{b103} an independently  by \cite{b104} who used a potential of the form $V(\Phi)=V_0[\mbox{cosh}(\eta\Phi-1)]$ where $V_0$ and $\eta$ are constants to explain the core density problem for disc galaxy halos in the $\Lambda$CDM model (see also \cite{b105,b106}) and to perform the fist cosmological analysis in the context of SFDM. They showed that the evolution of the Universe, its expansion rate and the growth of linear perturbations in this model are identical as those derived in the standard model. In \cite{b1} they model the dark matter in spiral galaxies, assuming dark matter as an arbitrary scalar field endowed with a scalar potential. 

Another scalar potential widely used to describe dark matter is $V(\Phi)=m^2\Phi^2/2$ \cite{b107,b22}. This potential is very interesting because it can mimic the cosmological evolution of the Universe predicted by the $\Lambda$CDM model. Also, it is known that an exponential-like scalar field potential fits very well the cosmological constraints due to the form of its solutions (see for example \cite{b105,b116,b106}). If the self-interaction of the SF is considered, we need to add a quartic term to the SF potential \cite{b11,b23,b108,b109,b110}, in this case the equation of state of the SF results to be that of a polytrope of index n=1 (see \cite{b19,b20,b21,b111,b98}).

Another interesting result is that the predicted density of neutrinos at the recombination epoch is in agreement with the observations of the Wilkinson Microwave Anisotropy Probe (WMAP). In the same direction, Rodr\'iguez-Montoya et al. \cite{b17} studied ultra-light bosons as dark matter in the Universe with the framework of kinetic theory, through the Boltzmann-Einstein equations, and they found that this kind of ultra-light particles is consistent with the acoustic peaks of the cosmic microwave background radiation if the boson mass is around $m\sim 10^{-22}$eV.

\cite{b115} pointed out that SFDM/BEC can explain the spatial separation of the dark matter from visible matter, as derived from X-ray maps and weak gravitational lensing, in the Bullet Cluster, see also \cite{b43}. 

Other works have used the bosonic dark matter model to explain the structure formation via high-resolution simulations. \cite{b14,b112} reviewed the key properties that may arise from the bosonic nature of SFDM models. On the other hand, several authors have numerically studied the formation, collapse and viralization of SFDM/BEC halos as well as the dynamics of the SFDM around black holes \cite{b31,b32,b33,b117,b39,b118,b119,b37}. In \cite{b102}, Alcubierre et al. found that the critical mass for collapse of the SF is of the order of a Milky Way-sized halo mass. This suggests that SFDM/BEC can be plausible candidate to dark matter in galactic halos. In addition, Lora et al.,\cite{b42}, studied, through N-body simulations, the dynamics of Ursa Minor dwarf galaxy and its stellar clump assuming a SFDM/BEC halo to establish constraints for the boson mass. Moreover, they introduced a dynamical friction analysis with the SFDM/BEC model to study the distribution of globular clusters in Fornax. An overall good agreement is found for the ultra-light mass $\sim 10^{-22}$eV of bosonic dark matter.

In this model the scalar particles with ultra-light mass are such that their wave properties avoid the cusp problem and reduce the high number of small satellites by the quantum uncertainty principle (\cite{b12,b15,b96,b44,b97}). Robles and Matos \cite{b97}, (see also \cite{b113,b100,b96}) showed that BEC dark matter halos fit very well high-resolution rotation curves of LSB galaxies, and that the constant density core in dark halos can be reproduced. Also, \cite{b95} showed how the SFDM/BEC paradigm is a good alternative to explain the common mass of the dark halos of dwarf spheroidal galaxies. Recently, Rindler-Daller and Shapiro, \cite{b40}, investigated the formation of vortex in SFDM/BEC halos. They found constraints on the boson mass in agreement with the ultra-light mass found in previous works (see also \cite{b41,b120}), some 
of these issues will be discussed in more detail in Section 4.

Summarizing, it is remarkable that with only one free parameter, the ultra-light scalar field mass ($m\sim 10^{-22}$eV), the SFDM model fits:

i) The evolution of the cosmological densities \cite{b22}.

ii) The central density profile of the dark matter is flat \cite{b100}.

iii) The acoustic peaks of the cosmic microwave background \cite{b17}.

iv) The scalar field has a natural cut off, thus the substructures in cluster galaxies are suppressed naturally. With a scalar field mass of $m_{\Phi}\sim 10^{-22}$eV the amount of substructures is compatible with the ones observed \cite{b10,b12,b19}.

v) We expect that SFDM forms galaxies earlier than the $\Lambda$CDM model, because they form BEC's at a critical temperature $T_c>>$MeV. So if the SFDM model is right, we have to see big galaxies at high redshifts with similar features \cite{b22,b19}. 

vi) Adding self-interaction and Temperature correcctions, the rotation curves of big galaxies and LSB galaxies \cite{b101,b98,b99, b21, b100}.

vii) With this mass, the critical mass of collapse for a real scalar field is just $10^{12}M_{\odot}$, i. e., the observed in galaxy halos \cite{b102}.

vii) The observed properties of dwarf galaxies, i. e., the minimum length scale, the minimum mass scale, and their independence from the brightness \cite{b95}.

ix) And recently it has been demonstrated that the SFDM halos would have cores large enough to explain the longevity of the cold clump in Ursa Minor and the wide distribution of globular clusters in Fornax \cite{b42}.

Then, the SFDM/BEC model has provided to be a good candidate for dark matter halos of galaxies in the Universe because it can explain many aspects where the standard model of cosmology fails (\cite{b100,b113,b114,b95,b115,b42,b96,b97}). Therefore, not only the many successful predictions of the Standard Model of cosmology at large scales are well reproduced by SFDM, but also 
the ones at galactic scales. The scalar field models presents some advantages over the standard $\Lambda$CDM model like the ones mentioned above. Also, its self-interaction can, in principle, explain the smoothness of the energy density profile in the core of galaxies \cite{b10,b125}. Nevertheless, its important to remark that
when a new dark matter candidate is proposed the study of the final object that will be formed as a result of a gravitational collapse is always 
an important but difficult task that requires continuos work since the baryonic physics is still poorly understood.  

\section{Current Status of the SFDM model}

\subsection{Self-gravitating Bose-Einstein condensate dark matter}

Following \cite{b113}, dark matter halos as a self-gravitating Bose-Einstein condensate with short-range interactions have been 
widely discussed \cite{b101,b97,b38,b39}. In these models, it is supposed that the cosmic BEC has a relatively low mean mass density so that the Newtonian approximation can be used. In the literature, when $T=0$ all the bosons have condensed and the system can be described by one order parameter $\psi(\vec{r},t)$, called the condensate wave function. In the mean-field approximation, the ground state properties of the condensate are described by the Gross-Pitaevskii equation:
$$i\hbar\frac{\partial\psi}{\partial t}(\vec{r},t)=-\frac{\hbar^2}{2m}\Delta\psi(\vec{r},t)+m\Phi_{tot}(\vec{r},t)\psi(\vec{r},t).$$
where $\Phi_{tot}$ is the total potential exerted on the condensate.
With this equation, \cite{b38} studied the structure and the stability of a self-gravitating BEC with short-range interactions. 
In this case, the results obtained in the absence of self-coupling and the results of \cite{b113} obtained for self-coupled BECs in the Thomas-Fermi approximation have been connected. The case of attractive short-range interactions where considered and the existence of a maximum mass above which no equilibrium state exists was found.

This study was motivated by the proposal that dark matter halos could be gigantic cosmic BEC's, \cite{b12,b5,b124,b99,b113}. In this case, gravitational collapse is prevented by the Heisenberg uncertainty principle or by the short-range interaction. This suggestion still remains highly speculative since the nature of dark matter remains unknown. On the other hand, whatever the nature of its constituents, if dark matter is viewed as a collisionless system described by the Vlasov equation, dark matter halos could result from processes of violent collisionless relaxation. In that case, gravitational collapse can be prevented by a Lynden-Bell�s type of exclusion principle, because this form of relaxation could result more rapid and efficient than a �collisional� relaxation. Furthermore, it generates a density profile with a flat core and a $r^{-2}$ outer density profile for the halo, yielding flat rotation curves. These features resulted remarkably consistent with observations making this alternative scenario quite attractive.

In \cite{b39} the same author obtained the exact mass-radius relation of self-gravitating BECs with short-range interactions by numerically solving the equation of hydrostatic equilibrium this time taking into account quantum effects. He compared his results with the approximate analytical relation obtained in \cite{b38} from a Gaussian ansatz. He found that the Gaussian ansatz always provided a good qualitative agreement with the exact solution, and that the agreement was quantitatively very good.

In one of his most recent works Chavanis \cite{b25} assumed that the dark matter in the universe could be a self-gravitating BEC with short-range interactions, and he then theoretically explored the consequences of this hypothesis. This time he considered the possibilities of positive and negative scattering lengths.

At the level of dark matter halos a positive scattering length, equivalent to a repulsive self-interaction generating a positive pressure, is able to stabilize the halos with respect to gravitational collapse. This leads to dark matter halos without density cusps with an effective equation of state
equal to that of a polytrope of index $n=1$. Alternatively, if the scattering length is negative, equivalent to an attractive self-interaction generating a negative pressure, the dark matter results very unstable and collapses above a very small critical mass $M_{max}=1.012\hbar/\sqrt{|a_s|Gm}$, where $a_s$ is the scattering length. When these ideas where applied to an infinite homogeneous cosmic fluid, it was found that a negative scattering length can increase the maximum growth rate of the instability and accelerate the formation of structures. The virtues of these results could be combined by assuming that the scattering length changes sign in the course of the evolution. It could be initially negative to help with the formation of structures and become positive to prevent complete gravitational collapse. The mechanism behind this change of sign remains unknown. However, some terrestrial experiments have demonstrated that certain atoms can have negative scattering lengths and that it is possible in principle to manipulate the value and the sign of $a_s$. 

It has also been found that a SFDM/BEC universe with positive scattering length, having a positive pressure, is not qualitatively very different from a classical Einstein-de Sitter universe. It also emerges at a primordial time $t=0$ from a big-bang singularity where the density results infinite, and then undergoes a decelerating expansion. A difference, however, is that the initial scale factor $a(0)$ is finite. On the other hand, a SFDM/BEC universe with an always negative scattering length, having a negative pressure, markedly differs from previous models. It starts from $t\longrightarrow -\infty$ with a vanishing radius and a finite density, it has an initial accelerating expansion then decelerates and asymptotically behaves like the EdS universe. This model universe exists for any time in the past and there is no big-bang singularity. When the effect of radiation, baryonic matter and dark energy where added, the picture resulted quite different. In that case, a SFDM/BEC universe with attractive or repulsive self-interaction started from a singularity at $t=0$ where the density was infinite. It first experiences a phase of decelerating expansion followed by a phase of accelerating expansion. For $k\longrightarrow 0$ ($k$ being the wave number) the standard $\Lambda$CDM model was recovered but for $k\neq0$, the evolution of the scale factor in a SFDM/BEC universe turned out substantially different. The model with $k>0$ expanded more rapidly than the standard model. The initial scale factor is finite and the radiation never dominates. The model with $k<0$ expanded less rapidly than the standard model. The initial scale factor vanishes and the radiation dominates leading to a decelerating expansion. In both models, the dark energy dominates at large times leading to an accelerating expansion. Finally, a �dark fluid� with generalized equation of state $p=(\alpha\rho+k\rho^2)c^2$ having a pressure component $p=k\rho^2c^2$ similar to a BEC dark matter and a component $p=\alpha\rho c^2$ mimicking the effect of the cosmological constant were considered. Optimal parameters ($\alpha$, $k$) that gave a good agreement with the standard model where found. Also the growth of perturbations in these different models where studied and confirmed the previous observation of Harko, \cite{b21}, that the density contrast increases more rapidly in a BEC universe than in the standard model.

In conclusion, it was pointed out that the idea that dark matter could be a BEC results fascinating and probably deserves further research.

\subsection{BEC dark matter and cosmological perturbations}

Through out this work we have shown how the SFDM/BEC model could be a serious alternative to the dark matter in the Universe. 
In \cite{b27} they studied in quite some detail the growth and virialization of $\Phi^2$-dark matter perturbations in the linear and nonlinear regimes. Following the spherical collapse model, they also studied the nonlinear regime of the evolution of $\Phi^2$-dark matter perturbations. They showed that the evolution of an overdense region of $\Phi^2$-dark matter can collapse and virialize in a bound structure. However, they found that the scalar perturbations collapse at earlier times of the Universe than those in the CDM model. Thus, the standard and the SFDM/BEC model can be confronted in their predictions concerning the formation of the first galaxies. Massive galaxies at high redshifts is a prediction of the model and may be used to distinguish between SFDM/BEC paradigm and CDM.

As the study in \cite{b27} is detail in the analysis of perturbations we include a brief summary. In the study of the cosmological dynamics of the SFDM model it was considered the simplest case: a single scalar field $\tilde\Phi(x,t)$,with self-interacting double-well potential. They used the potential 
$$V(\tilde\Phi)=\frac{\lambda}{4}\left(\tilde\Phi^2-\frac{\tilde m^2}{\lambda}\right)^2.$$
In a very early stage of the Universe, this scalar field was in local thermodynamic equilibrium with its surroundings see \cite{b110}. At some time, the scalar field decoupled from the rest of the matter and started a lonely journey with its temperature T decreased by the expansion of the Universe. Thus, it is considered the scalar field in a thermal bath of temperature T, whose scalar field potential, extended to one loop corrections, is given by
$$V(\tilde\Phi)=-\frac{1}{2}\tilde{m}^2\tilde\Phi^2+\frac{\lambda}{4}\tilde\Phi^4+\frac{\lambda}{8}T^2\tilde\Phi^2-\frac{\pi}{90}T^4+\frac{\tilde{m}^4}{4\lambda},$$ 
where $\tilde{m}$ is a mass parameter before the breaking of symmetry and $\lambda$ is the self-interacting constant. From here, it can be calculated the critical temperature $T_c$ at which the $Z_2$ symmetry of the real SF breaks. To do that, they calculated the critical points of the scalar potential from
$$0=\left(-\tilde{m}^2+\lambda\tilde\Phi^2+\frac{\lambda}{4}T^2\right)\tilde\Phi.$$
The negative term $-\tilde{m}^2$ permits the breaking of symmetry of the potential. One critical point is at $\tilde\Phi=0$. If the temperature T is high enough, the scalar potential has a minimum at this critical point. Furthermore, the critical temperature $T_c$ in which $\tilde\Phi=0$ becomes a maximum is
$$T_c^2=\frac{4\tilde{m}^2}{\lambda}.$$
This critical temperature defines the symmetry breaking scale of the scalar field.

To study the dynamics of the SFDM in the background Universe it is assumed a Friedmann-Lemaitre-Robertson-Walker metric with scale factor $a(t)$. The background Universe was composed by SFDM ($\Phi_0$) endowed with a scalar potential $V\equiv V(\Phi_0)$, baryons (b), radiation (z), neutrinos ($\nu$), and a cosmological constant ($\Lambda$) as dark energy. For the basic background equations, we have from the energy-momentum tensor $\boldsymbol{T}$ for a scalar field, the scalar energy density $T_0^0$ and the scalar pressure $T_j^i$ are given by
\begin{eqnarray}
T_0^0&=&-\rho_{\Phi_0}=-\left(\frac{1}{2}\dot\Phi_0^2+V\right),\\
T_j^i&=&P_{\Phi_0}=\left(\frac{1}{2}\dot\Phi_0^2-V\right)\delta^i_j,
\end{eqnarray}
where the dots stand for the derivative with respect to the cosmological time and $\delta_j^i$ is the Kronecker delta. Thus, the 
cosmological Equation of State for the scalar field is $P_{\Phi_0}=\omega_{\Phi_0}\rho_{\Phi_0}$ with
$$\omega_{\Phi_0}=\frac{\frac{1}{2}\dot\Phi_0^2-V}{\frac{1}{2}\dot\Phi_0^2+V}.$$
\begin{figure}[h]
\centering
 \scalebox{0.5}{\includegraphics{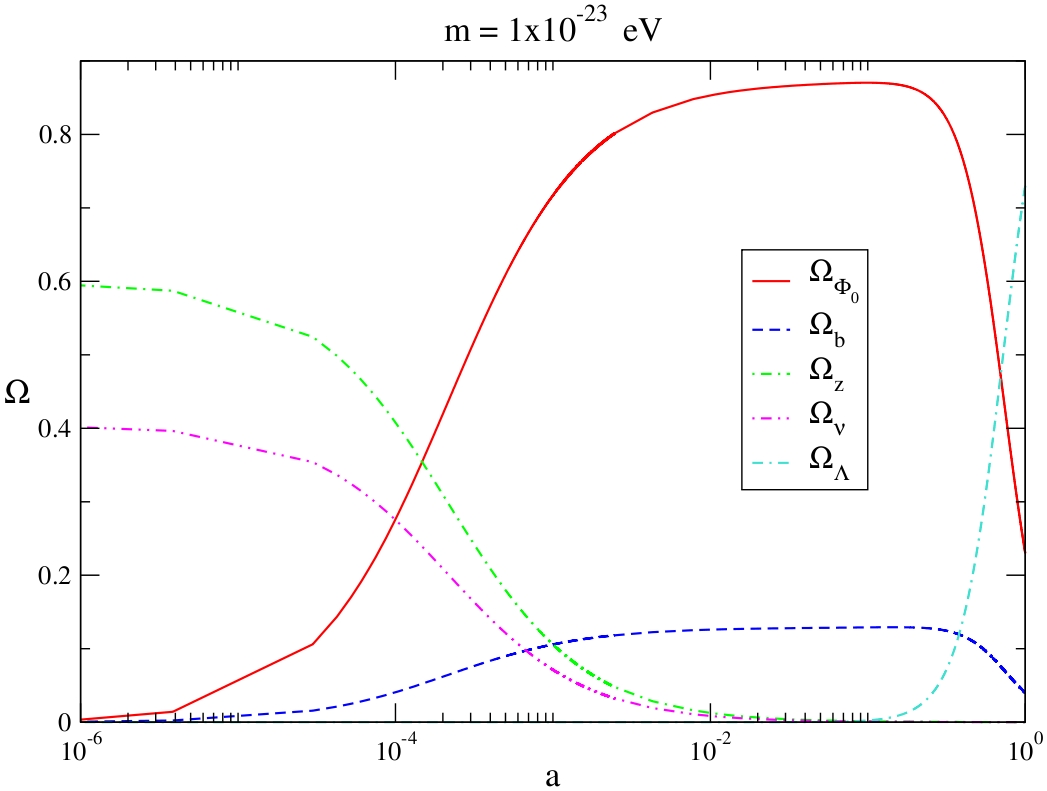}}
 \scalebox{0.5}{\includegraphics{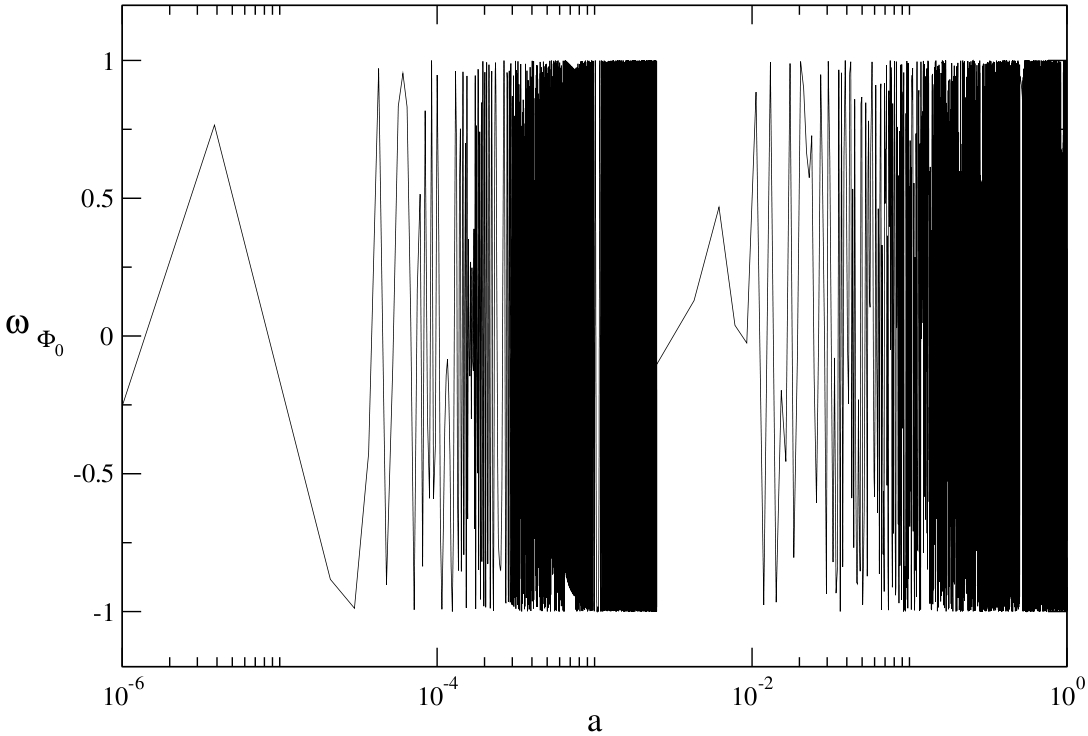}}
 \caption{Left: Evolution of the density parameters $\Omega_i$ for the background Universe. Scalar field dark matter model mimics the standard $\Lambda$CDM behavior. Right: Evolution of the scalar field dark matter equation of state for the background Universe.}
 \label{magana1}
\end{figure}

In order to solve the Friedmann equations with analytic methods with the approximation $m>>H$ they performed a transformation and compared their result with numerical ones. Here the scalar field and the variables of the background depend only on time, e.g., $\Phi=\Phi_0(t)$.

They computed the growth of the SFDM overdensities $\delta\rho_{\Phi}$ in the linear regime, in this regime, the density contrast $\delta\equiv\delta\rho_{\Phi}/\rho\Phi_0$ was much smaller than unity. It is believed that the Universe was almost uniform after inflation, with a very small density contrast. As the Universe expanded, the small overdensities grew until they began to collapse, leading to the formation of structure in the Universe. Thus, only small deviations in the FLRW model are considered, so that they can be treated by linear perturbation theory. After introducing the perturbed metric tensor in the FLRW background, only scalar perturbations considered. In their paper they gave the equation of energy-momentum conservation and the Einstein field equations for the perturbed metric.
Within the linear theory of scalar perturbations the evolution of the density contrast can be written as 
$$\dot\delta+3H\left(<\frac{\delta P_{\Phi}}{\delta\rho_{\Phi}}>-<w_{\Phi_0}>\right)\delta=3\dot\phi_k<F_{\Phi}>-<G_{\Phi}>$$
where
\begin{eqnarray}
F_{\Phi}&=&1+w_{\Phi_0}\nonumber\\
G_{\Phi}&=&\frac{2k^2}{a^2k^2}\frac{\dot\phi_k+H\phi_k}{\rho_{\phi_0}},
\end{eqnarray}
being $\phi_k$ the gravitational potential. This equation differs from the density contrast equation for $\Lambda$CDM. However, 
in \cite{b27} they show that the extra terms $F_{\Phi}$ and $G_{\Phi}$ tend to the values of the standard equation of $\Lambda$CDM.
Therefore the scalar perturbations in this model grow up exactly as in the $\Lambda$CDM paradigm.
\begin{figure}[h]
\centering
 \scalebox{0.65}{\includegraphics{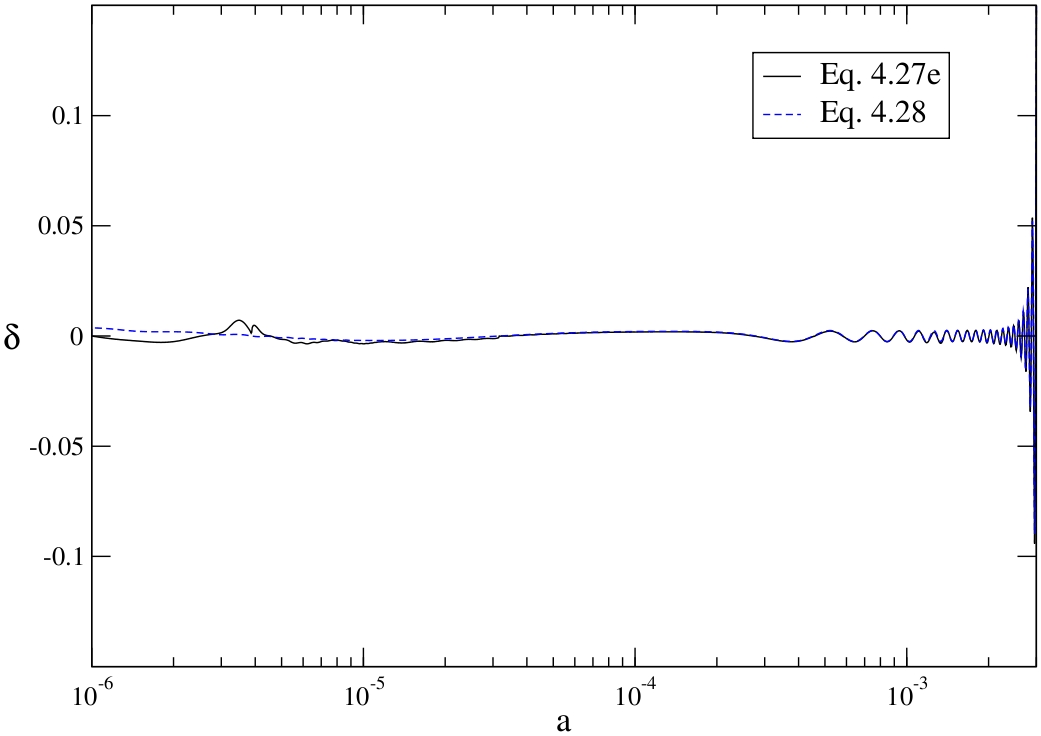}}
 \caption{Evolution of the density contrast $\delta$ for a perturbation with wavelength $\lambda_k\sim 2$Mpc}
 \label{magana2}
\end{figure}

The evolution of the scalar perturbations in the nonlinear regime when $\delta>>1$ was also studied. Here, an analysis was made 
within the framework of the spherical collapse model \cite{b52}. This formalism is very useful to understand the structure 
formation process in the Universe in the nonlinear regime. Focused on the era where the radiation density is equal to the 
SFDM density, the $T<<T_c$, and therefore,it is expected that the scalar potential reaches the $\Phi^2_0$ profile.
They also studied if $\Phi^{2}$-dark matter perturbations (once the breaking of symmetry was achieved and the SF had reached its minimum ) where able to form bound structures as in the standard model.

\begin{figure}[h]
\centering
 \scalebox{0.45}{\includegraphics{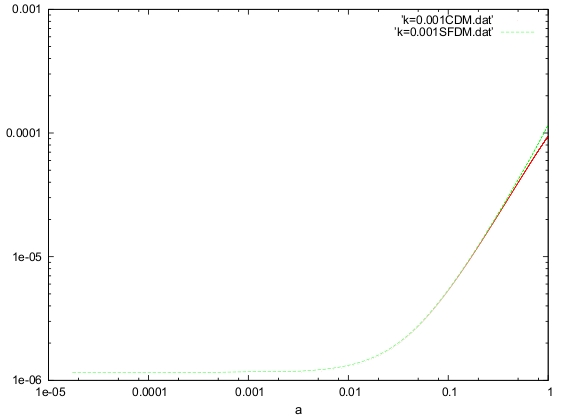}}
\caption{Evolution of the perturbations for the CDM model (red dots) and SFDM model (green line) for $k = 1\times10^{-3}hMpc^{-1}$. Notice how after the epoch of equality ($a_{eq}\sim 10^{-4}$) the evolution of both perturbations is identical, $a=1$ today. In this case the self-interacting parameter is $\lambda=0$\cite{b19}}
 \label{suarez1}
\end{figure}

Following the same path, \cite{b19} obtained that for the matter dominated era the low-$k$ modes grow. When CDM decouples from radiation in a time just before recombination it grows in a milder way than it does in the matter dominated era (Fig. \ref{suarez1}).
 
Although in general a scalar field is not a fluid, it can be treated as if it behaved like one and the evolution of its density can be the appropriate for the purpose of structure formation, because locations with a high density of dark matter can support the formation of galactic structure.

In \cite{b19} they assumed that there was only one component to the mass density, and that this component was given by the scalar field dark matter. In this case the equation for the perturbations reads
$$\frac{d^2\delta}{dt^2}+2H\frac{d\delta}{dt}+\left[(v_q^2+w\hat\rho_0)\frac{k^2}{a^2}-4\pi G\hat\rho_0\right]\delta=0,$$
valid for all sub-horizon sized perturbations in the non-relativistic regime.

\begin{figure}[h]
\centering
 \scalebox{0.45}{\includegraphics{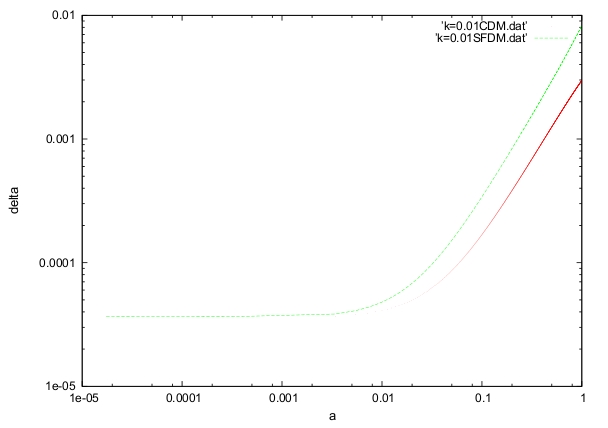}}
\caption{Evolution of the perturbations for the CDM model (red dots) and SFDM model (green line) for $k=1\times 10^{-2}hMpc^{-1}$ and $\lambda\neq$ and negative. Notice how after the epoch of equality ($a_eq\sim 10^{-4}$) the evolution of both perturbations is now different from the one in Fig. \ref{suarez1}, $a=1$ today. In this case we can clearly see that the SFDM fluctuations grow faster than those for the CDM model.}
 \label{suarez2}
\end{figure}

It was shown that the scalar field with an ultralight mass of $10^{-22}$eV simulates the behavior of CDM in a Universe dominated by matter when $\lambda=0$, because in general in a matter dominated Universe for low-$k$, $v_q$ (called the quantum velocity) tends to be a very small quantity tending to zero, so from the equation of the density contrast we could see that on this era we have the $\Lambda$CDM profile given by 
$$\frac{d^2\delta}{dt^2}+2H\frac{d\delta}{dt}+\left(c_s^2\frac{k^2}{a^2}-4\pi G\hat\rho_0\right)\delta=0,$$
i.e., the SFDM density contrast profile is very similar to that of the $\Lambda$CDM model, Fig. \ref{suarez1}. On the contrary for $\lambda\neq 0$ both models have different behavior as can be seen from Fig. \ref{suarez2},
results which show that linear fluctuations on the SFDM can grow, even at early times when the large-scale modes (small $k$) have entered the horizon just after $a_{eq}\sim 10^{-4}$, when it has decoupled from radiation, so the amplitudes of the density contrast start to grow faster than those for CDM around $a\sim 10^{-2}$. Here an important point is that although CDM can grow it does so in a hierarchical way, while from Fig. \ref{suarez2} we can see that SFDM can have bigger fluctuations just before the $\Lambda$CDM model does, i.e., it might be that no hierarchical model of structure formation is needed for SFDM, and it is expected that 
for the non-linear fluctuations the behavior will be quite the same as soon as the scalar field condensates, which could be in a very early epoch when the energy of the Universe was about $\sim$ TeV. These facts can be the crucial difference between both models.

Additionally, in \cite{b13} the growth of cosmological perturbations to the energy density of dark matter during matter domination was 
considered when dark matter is a scalar field that has undergone Bose-Einstein condensation. In this case, the inhomogeneities where 
considered within the framework of both Newtonian gravity, where the calculation and results resulted more transparent, and General 
Relativity. The direction taken was again in deriving analytical expressions, which where obtained in the small pressure limit. 
Throughout their work the results where again compared to those of the standard cosmology, where dark matter is assumed pressureless, 
using analytical expressions to showcase precise differences. They also find, compared to the standard cosmology, that Bose-Einstein condensate dark matter leads to a scale factor, gravitational potential and density contrast that again increases at a faster rate.

\subsection{Galaxies and BEC dark matter mass constraints}

Maga\~{n}a et al.\cite{b42} considered a model where ultra-light bosons are the main components of the dark halos of galaxies. The main goal of this work was to constrain the mass of the scalar particles. They constructed stable equilibrium configurations of SFDM in the Newtonian limit to model the DM halo in UMi. They studied two relevant cases of SFDM halos: with and without self-interaction.

Since galactic halos are well described as Newtonian systems, the work was done within the Newtonian limit. In this limit, The Einstein-Klein-Gordon equations for a complex scalar field $\Phi$ minimally coupled to gravity and endowed with a SF potential $V(\Phi)=m_{\phi}^2\Phi^2/2+\lambda\Phi^4/4,$ can be simplified to the Schr\"odinger-Poisson equations:
\begin{eqnarray}
i\hbar\partial_t\psi&=&-\frac{\hbar^2}{2m_{\phi}}\nabla^2\psi+Um_{\phi}\psi+\frac{\lambda}{2m_{\phi}}|\psi|^2\psi,\\
\nabla^2U&=&4\pi Gm_{\phi}^2\psi\psi^*,
\end{eqnarray}
where $m_{\phi}$ is the mass of the boson associated with the scalar field, $U$ is the gravitational potential produced by the DM density core, $\lambda$ is the self-interacting coupling constant, and the field $\psi$ is related to the relativistic field $\Phi$ through
$$\Phi=e^{-im_{\phi}c^2t/\hbar}\psi.$$
 
UMi is a diffuse dSph galaxy located at a distance of $69\pm4$kpc from the Milky Way center and has a luminosity of $L_V=3\times10^5L_{\odot}$. Its stellar population is very old with an age of 10-12Gyr. Dynamical studies suggest that UMi is a galaxy dominated by DM, with a mass-to-light ratio larger than 60$M_{\odot}/L_{\odot}$. Among the most puzzling observed properties of UMi is that it hosts a stellar clump, which is believed to be a dynamical fossil that survived because the underlying DM gravitational potential is close to harmonic. This condition is accomplished if the DM halo has a large core.

\begin{figure}[h]
\centering
  \scalebox{0.53}{\includegraphics{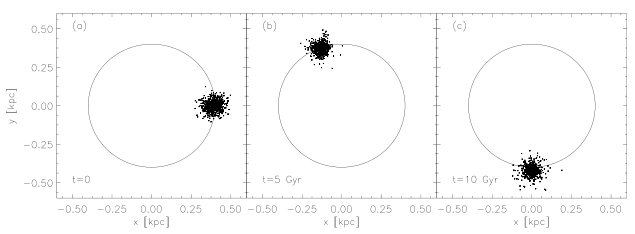}}
 \caption{Snapshots of the clump in UMi galaxy, at $t=0, 5$, and 10Gyr.  The clump is set on a circular orbit in the $(x,y)$-plane at a distance of $ =0.39$kpc from UMi�s center. The mass of the boson is $m_{\phi}=10^{-23}$eV and $\Lambda=0$. The total mass of the galaxy is $M=9.7\times 10^9M_{\odot}$.}
 \label{Lora}
\end{figure}

In that work, it is mentioned that the most remarkable feature in UMi structure is the double off-centered density peak. The second peak or clump is located on the north-eastern side of the major axis of UMi at a distance of $\sim$0.4kpc from UMi�s center. The velocity distribution of the stars contained in the clump is well fitted by two Gaussians, one representing the background. The most appealing interpretation is that UMi�s clump is a disrupted cluster with an orbit in the plane of the sky, which has survived in phase-space because the underlying gravitational potential is harmonic, implying that the dark halo in UMi has a large 
core \ref{Lora}.

The fact that SFDM halos have cores might solve other apparent problems in dwarf galaxies. In their work the timing problem of the 
orbit decay of GCs (Globular clusters) in dwarf elliptical galaxies and dSph galaxies was considered. In fact, in a cuspy halo, GCs in these galaxies 
would have suffered a rapid orbital decay to the center due to dynamical friction in one Hubble time, forming a nucleated dwarf galaxy. For instance, under the assumption that mass follows light or assuming a NFW profile, Fornax GCs 3 and 4, which are at distances to the 
center $<$0.6kpc, should have decayed to the center of Fornax in $\sim$ 0.5-1Gyr; this clearly represents a timing problem. Assuming a cuspy NFW halo, GCs 1, 2, 3 and 5 can remain in orbit as long as their starting distances from Fornax center are $\gtrsim$1.6kpc, whereas one of them needs an initial distance $\gtrsim$1.2kpc. However, there is no statistical evidence to suggest that the initial distribution of GCs is so different to the stellar background distribution. In addition, studies of the radial distribution of GCs in giant elliptical galaxies show that the distribution of metal-rich GCs matches the galaxy light distribution. Assuming that GCs formed along with the bulk of the field star population in dwarf galaxies, the probability that Fornax GCs were formed all beyond 1.2kpc is $\sim(0.03)^5\simeq 2.5\times 10^{-8}$. Therefore, it is very unlikely that all the GCs in Fornax were formed at such large distances and even if they did, there is still a timing problem.

The persistence of cold substructures in UMi places upper limits on $m_{\phi}$. Using N-body simulations, it was found that the survival of cold substructures in UMi was only possible if $m_\phi<3\times 10^{-22}$eV in the $\Lambda=0$ case. On the other hand, by imposing a plausible upper limit on $M$, lower limits on $m_\phi$ where placed. All together, it was found that for $\Lambda=0$, $m_\phi$ should be in the window
$$0.3\times 10^{-22}\mbox{eV}<m_{\phi}<3\times 10^{-22}\mbox{eV}.$$
Since the timing problem of the orbital decay of the GCs in Fornax can be alleviated if $m_{\phi}<1\times10^{-22}$ eV for $\Lambda=0$, the most favored value resulted around $(0.3-1)\times 10^{-22}$eV.

For SFDM models with self-interaction, the upper limit on $m_{\phi}$ increases with $\Lambda$. Bosons of mass $\lesssim 6\times 10^{-22}$eV could account for the observed internal dynamics of UMi.  In the limit $\Lambda>>1$, it was found that $m^4_{\phi}/\lambda\lesssim 0.55\times 10^3\mbox{eV}^4$ would explain the longevity of UMi�s clump and the surviving problem of GCs in Fornax.

The window of permitted values for $m_{\phi}$ resulted quite narrow. Even so, it is remarkable that the preferred range for the mass of the boson derived from the dynamics of dSph galaxies resulted compatible with those given by other authors to ameliorate the problem of overabundance of substructure and is also consistent with the CMB radiation \cite{b12,b103,b17}.

In a recent posting, Slepian and Goodman constrained the mass of bosonic DM using rotation curves of galaxies, 
and Bullet Cluster measurements of the scattering cross section of self-interacting DM under the assumption that 
these systems are in thermodynamic equilibrium. If their assumptions are verified, repulsive bosonic DM will be excluded and, thereby, the only remaining window open is non-interacting bosons. Nevertheless, the static diffusive equilibrium between Bose-Einstein condensate and its non-condensated envelope, as well as finite temperature effects need to be reconsidered. In addition, other authors \cite{b40,b18} argue that scattering cross sections for bosonic DM are much smaller than those derived from the condition of thermodynamic equilibrium by Slepian and Goodman.

\subsection{SFDM density profiles and LSB galaxies}

An analysis of the Newtonian regime at Temperature zero can be found in Bohmer and Harko (2007), they assumed 
the Thomas-Fermi approximation which neglects the anisotropic pressure terms that are relevant only in the boundary of the condensate, 
the system of equations describing the static BEC in a gravitational potential $V$ is given by
\begin{eqnarray}
\nabla p\left(\frac{p}{m}\right)&=&-p\nabla V,\\
\nabla^2V&=&4\pi G\rho,
\end{eqnarray}
with equation of state
$$p(\rho)=U_0\rho^2,$$
where $U_0=\frac{2\pi\hbar^2a}{m}$, $\rho$ is the mass density of the static BEC configuration and $p$ is the pressure at zero 
temperature, $p$ is not the usual thermal pressure but instead it is produced by the strong repulsive interaction between the ground state bosons.

In \cite{b97}, Robles and Matos found that the BEC dark matter model can give a density contrast profile consistent with RC's of dark 
matter dominated galaxies. The profile resulted as good as one of the most frequently used empirical core profiles, the pseudo Isothermal profile (PI),
but with the advantage of coming from a solid theoretical frame. In \cite{b97} The data was fitted within 1kpc and a logarithmic 
slope $\alpha=-0.27\pm 0.18$ was found in perfect agreement with a core. They emphasized that the cusp in the central regions is not a prediction that 
comes from first principles in the CDM model, it is a property that is derived by fitting simulations that use only DM. For a detail
discussion of the cusp and core problem in the standard model see \cite{deBlock} and references there in.
They also explained an ambiguity in the usual interpretation of the core radius, they proposed a new definition for the core and core 
radius that takes away this ambiguity and that has a clear meaning that allows for a definite distinction of when a density profile is core or cusp.
Using their definition they found the core radius in the BEC profile to be in most cases over 2kpc bigger than the core radius in the PI 
profile. They assumed a great number particles were in the ground state in the form of a condensate. 
This led to good results for their sample of galaxies, but it proved necessary to consider more than these simple hypotheses when dealing with 
large galaxies.

\begin{figure}[h]
\centering
\begin{tabular}{cc}
\includegraphics{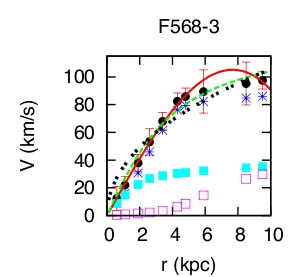} &
\includegraphics{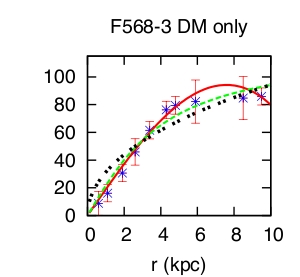} \\
\includegraphics{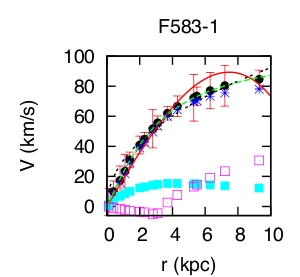} & 
\includegraphics{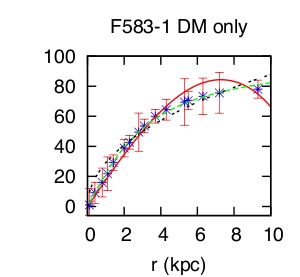} \\
\end{tabular}
\caption{Contribution of the baryons to the rotation curve for F568-3 and F583-1.
We denote observed data by black dots with error bars, dark matter with blue asterisks,
the disk with cyan squares and the the gas with magenta squared boxes. The figures on the left were fitted asumming the minimum disk hypothesis while the ones on the right are only the dark matter. In the fits shown are, BEC in solid line(red in 
the online version), PI dashed (green in the online version) and NFW double-dashed (black in the online version) profiles.}
 \label{robles1}
 \end{figure}

The solution to the system above is
\begin{equation}
\rho_B(r)=\rho_0^B\frac{sin(kr)}{kr} 
\label{densitycero}
\end{equation}
it can be seen that the BEC model satisfies $\rho\sim r^0$ near the origin, but a priori this does not imply consistency with 
observed RC's. Therefore, the profiles were fitted to thirteen high resolution RC data of a sample of LSB galaxies. 
The RC's were taken from a subsample of de Blok et al. (2001), galaxies that have at least 3 values within $\sim$1kpc where chosen, 
not presenting bulbs and the quality in the RC in $H_{\alpha}$ is good as defined in McGaugh S. S. et al. (2001). 
Because the DM is the dominant mass component for these galaxies they adopt 
the minimum disk hypothesis which neglects baryon contribution to the observed RC. In order to show that in LSB and dwarf 
galaxies neglecting the effect of baryons was a good hypothesis, they included two representative examples, see Fig. \ref{robles1}.

For these galaxies the contribution of the gas was plotted, disc and dark matter separately. They did the fitting first considering 
the total contribution and then using only DM. They found no substantial difference in their values.

As a second result and direct consequence of the core definition, they where able to obtain the constant value of $\mu_0$ which is proportional to the central surface density. This result is one of the conflicts of the current standard cosmological model
due to the hierarchical formation of galaxies.

As the density profile eq.(\ref{densitycero}) is not enogh to describe the large galaxies as discussed in \cite{b101}, 
Robles and Matos thus gave a physically motivated extension to the SFDM model that includes the DM temperature corrections 
to the first loop in perturbations. Their idea is to use the $Z_2$ spontaneous symmetry break of a real scalar field as a new 
mechanism in which the early DM halos form. As stated earlier, when the real scalar field rolls down to the minimum of the potential, 
the perturbations of the field can form and grow. They gave an exact analytic solution for an static spherically symmetric SF 
configuration, which in the SFDM model represents a DM halo. Their solution naturally presents a flat central density profile,just as
eq.(\ref{densitycero}), but now it can accommodate more than just the ground state as the temperature $T\neq 0$, in this way they  
solved previous discrepancies in rotation curve fits at $T=0$, for instance, having a constant halo radius for all galaxies and
the incapability to fit at the same time the inner and outermost regions of RC in large galaxies. Both issues were solved using
this scenario which includes temperature of the DM and the exited states of the SF.

The perturbed system of a scalar field with a quartic repulsive interaction but with temperature zero has been studied before \cite{b34,b30}. 
The study for the evolution of the SF with the temperature correction in a FRW universe is analogous. 
The metric tensor was written as $\boldsymbol{g}=\boldsymbol{g}_0+\delta\boldsymbol{g}$, where as always $\boldsymbol{g}_0$ is the unperturbed FRW background metric and $\delta\boldsymbol{g}$ the perturbation. The perturbed line element in conformal time $\eta$ given by
$$ds^2=a(\eta)^2[-(1+2\psi)d\eta^2+2B,_id\eta dx^i+(1-2\phi)\delta_{ij}+2E,_{ij}dx^idx^j]$$
with $a$ the scale factor, $\psi$ the lapse function, $\phi$ gravitational potential, B the shift, and E the anisotropic potential. The energy-momemtum tensor and the field where separated as $\boldsymbol{T}= \boldsymbol{T}_0+\delta\boldsymbol{T}$ and $\Phi(x^{\mu})=\Phi_0(\eta)+\delta\Phi(x^{\mu})$ respectively. As the linear regime was studied $\delta\Phi(x^{\mu})<<\Phi_0(\eta)$, the approximation $V(\Phi)\approx V(\Phi_0)$ could be made. They worked in the Newtonian gauge where the metric tensor $\boldsymbol{g}$ becomes diagonal and as a result, in the trace of the Einstein�s equations the scalar potentials $\psi$ and $\phi$ are identical, therefore, $\psi$ relates to the gravitational potential.

The work was mainly focused in the galactic scale DM halos after their formation. Robles and Matos constrained themselves to solve 
the Newtonian limit of the perturbed KG equation, that is,
$$\Box\delta\Phi+\frac{\hat\lambda}{4}[k_B^2(T^2-T_C^2)+12\Phi_0^2]\delta\Phi-4\dot\Phi_0\dot\phi+\frac{\hat\lambda}{2}[k_B^2(T^2-T_C^2)+4\Phi_0^2]\Phi_0\phi=0$$
valid when $\Phi$ is near the minimum of the potential and after the SB, where the SF is expected to be stable.Here $T_C$ is the 
temperature of the symmetry break.

\begin{figure}[h]
\centering
\begin{tabular}{cc}
\includegraphics{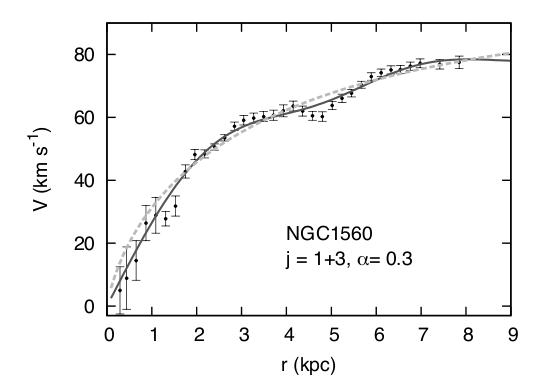} &
\includegraphics{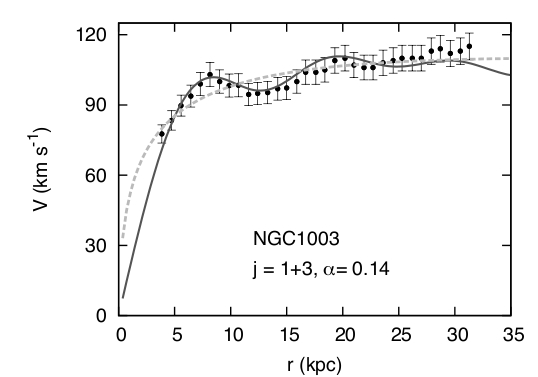} \\
\end{tabular}
\caption{Rotation curve of NGC 1560 and NGC1003 from \cite{b101}, where non zero temperature is considered. The wiggles are visible in
the outer region of the RCs.The solid line is the temperature corrected velocity profile and the dashed line is the Einasto fit\cite{b101}.}
 \label{rctemp}
 \end{figure}

Additionally to solving these two disagreements they mentioned why it does not seem necessary to include high amounts of feedback to 
fit and reproduce the inner core and wiggles found in high-resolution RC's,see Fig. \ref{rctemp}. Also, this model can be tested with 
high redshift observations, the SFDM model predicts initial core profiles as opposed to the initially cuspy ones found in CDM simulatios 
which are expected to flatten due to redistribution of DM by astrophysical processes. 

Finally one conclusion is that if observations of more galaxies with core behavior are confirmed, this model can be a good alternative 
to $\Lambda$CDM. 

\subsection{BEC dark matter and the power spectrum}

From another point of view in \cite{b26} it was again assumed that dark matter is composed of scalar particles that are able to form 
a Bose-Einstein condensate at some critical redshift $z_{cr}$, but in this case it was used to study the matter power spectrum.

After the BEC forms its effective pressure can assume a polytropic equation of state such as $P_{be}\sim\rho_{be}^{\gamma}$ if an 
arbitrary non-linearity term is assumed, in this case $\gamma$. The exact value of $\gamma$ is defined by the non-linear contribution of 
the Gross-Pitaevskii equation which in its standard form leads to \cite{b113}
$$\rho_{be}=\frac{2\pi\hbar^2l_a}{m_{\chi}^3}\rho_{be}^2.$$
In this case, the scattering length $l_a$ and the mass $m_{\chi}$ of the dark matter particle again determine the dynamics of the fluid. Assuming that the condensate does not interact with any other form of energy, the above pressure, via the conservation balance, leads to
$$\rho_{be}=\frac{m_{\chi}^3}{2\pi\hbar^2l_a}\frac{\rho_0}{a^3-\rho_0}$$
where
$$\rho_0=\frac{1.266\times\Omega_{be0}\times(m_{\chi}/1\mbox{meV})^{-3}\times(l_a/10^9\mbox{fm})}{1+1.266\times\Omega_{be0}\times(m_{\chi}/1\mbox{meV})^{-3}\times(l_a/10^9\mbox{fm})}.$$
The current value of the scale factor $a$ was taken as $a_0=1$ and the current fractional density of the BEC dark matter as $\Omega_{be0}=\rho_{be0}/\rho_c$ where $\rho_c$ is the critical density. When the above relations are combined the equation of state parameter of the BEC dark matter is obtained 
$$w_{be}=\frac{\rho_0}{a^3-\rho_0}.$$

If the inertial effects of the pressure become relevant as for example during the radiation phase or at the onset of the accelerated expansion, Newtonian cosmology fails and a more appropriate set of equations is needed. The inclusion of pressure in the Newtonian cosmology gives rise to the neo-Newtonian cosmology.
In this case, the matter power spectrum is defined as always
$$P(k)=|\delta_b(z=0;k)|^2,$$
where $\delta_{b}(k)$ is the baryonic density contrast calculated from equations
$$\delta_b''+\delta_b'\left(\frac{H'}{H}+\frac{3}{a}\right)-\frac{3}{2}\frac{\Omega_b}{H^2a^2}\delta_b=\frac{3\Omega_{be}}{2H^2a^2}(1+c_s^2)\delta_{be},$$
and
\begin{eqnarray}
\delta_{be}''&+&\left(\frac{H'}{H}+\frac{3}{a}-\frac{w_{be}'}{1+w_{be}}-\frac{3w_{be}}{a}\right)\delta_{be}'\nonumber\\
&+&\left[3w_{be}\left[\frac{H'}{Ha}+\frac{(2-3w_{be})}{a^2}\right]+\frac{3w_{be}'}{a(1+w_{be})+\frac{(k/k_0)^2c_s^2}{H^2a^4}}-\frac{3}{2}\frac{\Omega_{be}}{H^2a^2}(1+3c_s^2)(1+w_{be})\right]\delta_{be}\nonumber\\
&=&\frac{3}{2}\frac{\Omega_b}{H^2a^2}(1+c_s^2)\delta_b
\end{eqnarray}
at the present time \cite{b26}. The baryonic agglomeration $\delta_b$ is supposed to be driven by the gravitational field which is sourced by all the forms of energy. In order to solve this set of equations it was needed to set the initial conditions for $\delta_b$ and $\delta_{be}$ and their derivatives at $z_{cr}$ where the condensation took place. Since $z_{cr}=z_{cr}(l_a,m_{\chi})$ for each chosen couple of values $(l_a,m_{\chi})$ different initial conditions where needed.

In \cite{b26} it was shown that if such phase transition occurred in the recent Universe this process would be able to leave small, 
but perceptible, imprints on the large scale structure perceptible in the matter power spectrum. Assuming $l_a=10^6$fm the BEC 
dark matter model does shows differences of the order of a few percents for masses 15-35 meV. Adopting $l_a=10^{10}$fm corrections of the same order where obtained for masses 300-700meV.

Although the BEC phase is shown to have a small influence on the matter power spectrum, a more quantitative analysis could be performed to estimate the preferred values of the model parameters.

For the relevant parameter values studied in that work, the transition to the BEC phase was shown to occur at low redshifts. Since the standard cosmology remains unchanged before $z_{cr}$ the CMB physics at the last scattering surface remained the same. However, the BEC dark matter would modify the gravitational potential just after $z_{cr}$ while the speed of sound is nonzero leading to a contribution to the integrated Sachs-Wolfe effect.

\subsection{Vortices in BEC dark matter}

The conditions to form vortices in a SFDM/BEC halo have also been studied in \cite{b40, b120}. As it has been pointed out, these halos 
can be described as fluids, obeying quantum-mechanical fluid equations, so that the effects that make up this form of dark matter behave differently from standard $\Lambda$CDM, resulting in new effects with potentially observable consequences. The idea is that ultralight particles with $m<<1$eV will have very large de Broglie wave lengths which means that quantum statistical effects are important and macroscopic coherent lumps of matter can emerge. These light Bose particles will have a transition temperature to the condensed state that is of order $T_c\sim 2\cdot 10^9$K, which is the expected temperature in the Universe after about 1 second. 

There are essentially two limiting cases that can be considered. First, for quantum-coherence to be relevant on the scale of a halo of radius $R$, the particle de-Broglie wavelength
$$\lambda_{deB}=\frac{h}{mv},$$ 
should be considered to be of the order of the halo size, $\lambda_{deB}\lesssim R$, or else require $\lambda_{deB}<<R$ but with a 
strong repulsive self-interaction to hold the halo up against gravity. In the first case, if $v\simeq v_{virial}$ for the halo, 
this translates into a condition for the dark matter particle mass, $m\gtrsim m_H=1.066\cdot 10^{-25}(R/100\mbox{kpc})^{-1/2}(M/10^{12}M_{\odot})^{-1/2}\mbox{eV cm}^3$, where $m_H$ is a mass that depends on the properties of the halo and $g_H$ is determined by the density and the radius of the halo. In the second case, for the repulsive 
self-interaction pressure force to exceed the quantum pressure, it is required that $g>>g_H=2.252\cdot 10^{-64}(R/100\mbox{kpc})(M/10^{12}M_{\odot})^{-1}\mbox{eV cm}^3$. If $R$ is taken to be the radius of the virialized object supported against gravity by the dominant repulsive self-interaction, this imposes a condition on the particle mass given by $m\gtrsim\frac{m_H}{4}\sqrt{15g/g_H}$.

However, it seems that rotating BEC haloes add new phenomenology, and the possibility to distinguish this form of dark matter from other candidates. To this aim, in \cite{b40} they have studied the question of whether an angular velocity could be sufficient to create vortices in BEC/CDM cosmologies. As quantum fluid systems, BEC haloes can be modeled as uniformly rotating ellipsoids, with and without internal motions superposed. To this aim, in \cite{b40} the authors derived equations which relate the eccentricities of haloes to their $\lambda$-spin parameter. Once the latter is fixed, the eccentricities can be uniquely determined. They analytically studied necessary and sufficient conditions for vortex formation. In their results they found that vortex formation requires as a necessary condition that the halo angular momentum satisfies $L\geq L_{QM}=N\hbar$, which implies a lower bound on $m/m_H$, i.e. on the dark matter particle mass. However, a sufficient condition for vortex formation could be established by an energy analysis, which aimed to find the conditions of when a vortex becomes energetically favored.

They studied two classes of models for rotating halos in order to analyze stability with respect to vortex formation in two limits, 
one for $L/L_{QM}>>1$ and for $L/L_{QM}=1$, respectively. In what they called Halo-Model A ($L/L_{QM}>>1$) these where modeled as homogeneous Maclaurin 
spheroids. The minimum angular momenta for vortex formation in this case was $(L/L_{QM})_{crit}=(5.65, 4.53, 4.02)$ for 
$\lambda=(0.01, 0.05, 0.1)$, respectively, which corresponded to a constraint on the particle mass $m/m_H\geq(m/m_H)_{crit}$, 
where $(m/m_H)_{crit}=(309.41, 49.52, 21.73)$, respectively. As long as $m/m_H$ satisfied this condition, the strength of the 
self-interaction also satisfied the condition $g/g_H\geq(g/g_H)_{crit}$, where $(g/g_H)_{crit}=(1.02\cdot 10^5,2549.24,454.54)$ for the same $\lambda$-values, respectively.

For Halo-Model B ($L/L_{QM}=1$), which was an $(n=1)$-polytropic Riemann-S ellipsoid, strictly irrotational prior to vortex formation, even $L/L_{QM}=1$ resulted to be sufficient for vortex formation if the self-interaction strength was large enough. The condition $L/L_{QM}=1$ fixed the value of $m/m_H$ for each $\lambda$ according to
\begin{eqnarray}
\frac{L}{L_{QM}}&=&\frac{m}{m_H}\frac{\kappa_n}{10}\frac{2\tilde\Omega\sqrt{1-e_1^2}e_1^4}{(2-e_1^2)(1-e_1^2)^{5/6}(1-e_2^2)^{1/3}}\nonumber\\
&=&\frac{m}{m_H}\frac{\kappa_n}{10}\times\left(\frac{2B_{12}}{q_n}\right)^{1/2}\left(2+\frac{e_1^4}{4(1-e_1^2)}\right)^{-1/2}\frac{e_1^4}{(1-e_1^2)^{5/6}(1-e_2^2)^{1/3}},
\end{eqnarray}
and the condition of virial equilibrium
$$y(x)=\frac{\pi}{\sqrt{8}}g(e_1,e_2)^{-1/2}x,$$
thereby also fixing $g/g_H$. For $\lambda=(0.01, 0.05, 0.1)$, these values where given by $m/m_H=(44.58, 9.49, 5.01)$ and $g/g_H=(1595.07, 68.00, 17.20)$, respectively. Halo-Model B then made vortex formation energetically favorable for these values of $m/m_H$ and $g/g_H$. They interpreted this to mean that, for $L/L_{QM}>1$,vortex formation will also be favored, as long as $g/g_H>(g/g_H)$. Furthermore, any values of $m/m_H$ and $g/g_H$ which satisfy the condition for vortex formation in Halo-Model A would automatically satisfy that found by Halo-Model B, which resulted less stringent but more accurate.

In conclusion they imagined vortex formation in BEC haloes composed of repulsively interacting particles as follows: If the angular momentum of a rotating BEC halo fulfills $L<L_{QM}$, no vortex would form, and the halo can be modeled by a mildly compressible, irrotational Riemann-S ellipsoid, which has a polytropic index of $n=1$. For $L=L_{QM}$, the irrotational Riemann-S ellipsoidal halo can make a transition to a non-rotating, spherical halo with a vortex at the center if the self-interaction is strong enough. For a range of angular momenta fulfilling $L_{QM}< L\leq 2L_{QM}$, a central vortex can be expected but now with the excess angular momentum deforming the halo such that again a Riemann-S ellipsoid forms. Finally, if $L>>L_{QM}$, oblate haloes described as Maclaurin spheroids had a central vortex if $m/m_H\geq(m/m_H)_{crit}$ and $g/g_H\geq(g/g_H)_{crit}$ with the critical values given by Halo-Model A. Those critical values determined when a single vortex was energetically favored, but since $L/L_{QM}>>1$, it is also possible that multiple vortices could form. 

From another point of view, in \cite{b120}  it was assumed that the particles where non-interacting and therefore only gravity acted on the system. Following \cite{b12}, the authors based on Jeans instability analysis to estimate their parameters. The growing mode under gravity was given by $e^{\gamma t}$ with $\gamma^2=4\pi G\rho$, whereas the free field was supposed oscillatory; $e^{-iEt}$ with $E=k^2/2m$. The latter was then written as $e^{\gamma t}$ with $\gamma^2=-(k^2/2m)^2$. Noting that this is like normal Jeans analysis with sound speed $c^2_s=k/2m$ then $\gamma^2=4\pi G\rho-(k^2/2m)^2$. Setting this to zero, for the Jeans scale they found
$$r_J=2\pi/k_J=\pi^{3/4}(G\rho)^{1/4}m^{-1/2}=55m_{22}^{-1/2}(\rho/\rho_b)^{-1/4}(\Omega_mh^2)^{-1/4}\mbox{kpc},$$
where $m_{22}=m/10^{-22}$eV, and the background density is $\rho_b=2.8\cdot 10^{11}\Omega_mh^2M_{\odot}\mbox{Mpc}^{-3}$. It is supposed that below the Jeans scale the perturbations are stable and above it they behave as ordinary CDM \cite{b12}. As always, the stability below the Jeans scale was guaranteed by the uncertainty principle. If the particles are confined further, their momenta increases and oppose gravitational contraction.

Moreover, in \cite{b120} the author also considered the suggestion that superfluid BEC dark matter in rotation would likely lead to vortices as seen in atomic BEC experiments.  As already noted in \cite{b124}, BEC dark matter with self-interactions could actually constitute a superfluid. In the case of a repulsive self-interaction in \cite{b120} the author too argued that the vortex size should be determined locally by the coherence length. This means that there  would be two scales in the problem: A galactic one, given by the de Broglie wavelength from the tiny mass, and a sub galactic one determined by the mass and the two-body interaction strength. He also explored the consequences of self-interactions on the virialization of gravitationally bound structures and found almost no effects for reasonable values of $m$ and $a$. Here the case considered was also that of a quartic self-interaction $\lambda\phi^4$. 

In \cite{b113} there was a brief discussion of the effect on the Lane-Emden equation, whereas in other works a BEC of axions with a single vortex arising from global rotation in the early Universe has been considered. This latter scenario is, however, less likely to occur since the global rotation rate of the Universe can be estimated from various observations and is very small but nevertheless non-zero.

Under the assumption of dark matter being an ultra-light BEC, the rotation of spiral galaxies would cause vortex lattices to form. 
In \cite{b120}, the author also considered possible effects of sub galactic vortices in the dark matter on the rotation velocity 
curves of virialized galaxies with standard dark matter halo profiles. He found that one can actually get substructure in the rotation curves that resemble some observations, but that this required large vortex core size and small vortex-vortex distances. The mass and interaction strength needed to realize this where found to be fine-tuned, but could possibly be accommodated in more general setups.

If dark matter contains a component of condensed BEC particles that is superfluid and if the halos are rotating then it is not inconceivable that there can be vortex formation. However, the quantized vortex discussion makes an important assumption about the coherence length, $\xi$, 
entering $\Omega_c$ (angular velocity of the halo) in 
$$\Omega_c=\frac{\hbar}{mR^2}\mbox{ln}\left(\frac{1.46R}{\xi}\right)=\frac{6.21\cdot 10^{-17}}{m_{22}R^2}\mbox{ln}\left(\frac{1.46R}{\xi}\right).$$
Here $\xi$ is taken to be of kpc size. It then results that with no self-interaction there is only the gravitational scale $\hbar^2/GMm^2$ available, which becomes of galactic size for masses $m\sim 10^{-22}$eV. However, when including self-interactions through the scattering length $a$, there is also a scale given by $\xi=1/\sqrt{8\pi an}$, which is the usual Gross-Pitaevskii coherence length. The latter coincided with the characteristic length over which the density is expect to go to zero in a vortex.  

If the additional assumption that the vacuum expectation value, $\phi_0$, arises from a mechanism that preserves parity, the interaction term results
$$\frac{(mc^2)^2\phi^4}{4(\hbar c)^2\phi_0^2}=g\frac{\phi^4}{4}.$$
This terms is of course merely the standard interaction term in the Gross-Pitaevskii theory of interacting condensed bosons. Therefore he also concludes that the self-interacting scenario emerges from this procedure.

These investigations and simple numerical experiments pointed to an interesting effect that could arise from bosonic dark matter. 
However, to fully explore the influence that vortex lattice formation and stellar feedback on structure formation has in luminous matter, an ultralight BEC dark matter component in large N-body simulations should be considered.

\subsection{Bose dark matter additional constraint}

In \cite{b18} BE condensation inside the primeval fireball, at zero-order in perturbation theory has been studied. Here, the process of condensation was considered to be driven by self-interactions of high-energy bosonic particles.

In this work it was found that in the instantaneous decoupling approximation, the subsequent evolution of the full bosonic system was only affected by the expansion rate of the Universe and small gravitational instabilities.

The evolution of bosonic DM after decoupling was analyzed as follows: their velocity and temperature affected only by the expansion rate of the Universe.

In
$$m=\frac{\Omega_c\rho_{cr}}{n_c^{(0)}}=\frac{\Omega_H\rho_{cr}}{n_T^{(0)}},$$
the temperature was needed in order to calculate the mass of the bosonic DM.  Here $\Omega_c$ and $\Omega_H$ represent the content of bosonic CDM and bosonic HDM (hot dark matter) respectively, $n_c$ is the number density of condensed bosons and $n_T$ the number density of thermal bosons.

As an additional remark, they also address a bound on the strength of the bosons self-interactions.

In their study, the bosonic DM parameters where addressed as the mass, $m$, and the factor $g_x$ (amount of degrees of freedom); where bounds on 
their values have been obtained from a statistical analysis of cosmological data. The constraints found for the temperature of the 
boson gas $T_0^{\phi}=2.14\pm 0.02$K, and for the boson-antiboson gas $T^{\phi\tilde\phi}_0=1.91\pm 0.05$K.

Finally, from a similar analysis for fermionis they found bounds in the sum of neutrino masses and the number of extra relativistic species, $\sum m_{\nu}\lesssim 0.45$eV, $N=1.10\pm 0.18$, in concordance with some previous reports.

In summary, they presented a generic study of DM based on BE condensation, from which, CDM and HDM, result intrinsically related.

\subsection{SFDM and black holes}

The rapid decay of the energy density of the scalar field for the case of super-massive black holes, indicates that scalar fields may not be maintained around a black hole during cosmological time scales in the whole space; so either the scalar field gets accreted or it escapes through future null infinity. The fact is that when a Schwarzschild black hole that is asymptotically flat is considered, scalar fields tend to vanish from the spatial domain.

An appropriate coordinate system to study the propagation of scalar fields in the Schwarzschild space-time is using hyperboloidal slices, because it has been seen that such slices reach future null infinity instead of spatial infinity, which results as a natural boundary for a wave-like processes, including electromagnetic fields and gravitational radiation.

In this case the Klein-Gordon equation for a scalar field $\tilde\phi_T$ can be written as \cite{b119}: 
$\tilde\Box\tilde\phi_T-\frac{d\tilde V}{d\phi_T}=0$, where $\tilde\Box\tilde\phi_T=\frac{1}{\sqrt{-\tilde g}}\partial_{\mu}[\sqrt{-\tilde g}\tilde g^{\mu\nu}\partial_{\nu}\tilde\phi_T]$, with a potential of the form $\tilde V=\frac{1}{2}m_B^2|\tilde\phi_T|^2+\frac{\lambda}{4}|\tilde\phi_T|^4$, where $m_B$ has the units of mass. Then the KG equation in the conformal metric can be expressed as:
$$\tilde\Box\tilde\phi_T-\frac{1}{6}\tilde R\tilde\phi_T-(m_B^2\tilde\phi_T+\lambda\tilde\phi_T^3)=\Omega^3\left[\Box\phi_T-\frac{1}{6}R\phi_T-(m_B^2\Omega^{-2}\phi_T+\lambda\phi_T^3)\right]=0,$$
provided the relationship between the physical scalar field $\tilde\phi_T$ and the conformal scalar field 
$\phi_T$ to be $\phi_T=\tilde\phi_T/\Omega$. Here $R=\frac{12\Omega}{r^2}[r+(2r-1)]$ is the Ricci scalar of the conformal metric and 
$\Box=\nabla^{\mu}\nabla_{\mu}$ corresponds to the conformal metric. The case for which this last equation results conformally invariant corresponds to the zero mass case $m_B=0$. 

In their work \cite{b119}, used initial scalar field profiles with $m_B^2=0.0,0.1,0.2$. In order to explore the parameter space, various values of the amplitude $A=0.01,0.1,1.0$ and different widths of a Gaussian pulse, which in physical units corresponded to $\sigma_1=0.5$, $\sigma_2=1$ and $\sigma_3=5$ to the right from $r_0=0.8$. This range of parameters allowed the authors to cover length ranges that involve Compton wave-length related to effects of interaction and reflection that may involve reflection and absorption effects.  

In their work, \cite{b119}, one of the conclusions was that one potential ingredient that would help at maintaining massive scalar field densities during longer times is the rotation of the black hole and also of the scalar field. It would then be of major interest to use foliations that approach future null infinity and study if the same effects occured and also the study of scalar field configurations with non-zero angular momentum. In fact, the results in their work where obtained assuming the maximum cross section of accretion due to the spherical symmetry, which in turn worked as upper bounds accretion rates in more general cases.

Yet another possibility results on considering solutions that asymptotically may contain a cosmological constant, which would be appropriate if a background energy density in the universe where assumed. This would imply that black hole candidates should not be considered to be asymptotically flat. Other possibilities may include black hole candidates of a different nature like boson stars.
 
\subsection{Threats to scalar field dark matter, black holes?}

As mentioned in the last subsection, the existence of a long-lasting scalar field configurations surrounding a black hole have 
been studied \cite{b119,b37}. Another motivation for these studies is the possibility that super-massive black holes at galactic centers may represent a serious threat to the scalar field dark matter models.

As a first step, a relatively simple model has been considered \cite{b37}. There, stationary scalar field configurations to the Klein-Gordon equation 
$$(\Box-\mu^2)\phi=0,$$
 where looked for on a Schwarzschild space-time background,
 $$ds^2=-N(r)dt^2+\frac{dr^2}{N(r)}+r^2d\Omega^2,\mbox{  }N(r):=1-2M/r,$$
 with the d'Alambertian operator defined as $\Box :=(1/\sqrt{-g})\partial_{\mu}(\sqrt{-g}g^{\mu\nu}\partial_{\nu})$, $M$ being the mass of the black hole and $d\Omega^2:=d\theta^2+\mbox{sin}^2\theta d\varphi^2$ the standard solid angle element. So far, the case has been restricted to the case of a canonical, massive, non self-interacting minimally coupled scalar field $\phi$. With these conventions $\phi$ results dimensionless, while $\mu$ has dimensions of $\mbox{length}^{-1}$. The associated quantum mechanical "mass" of the scalar field given by $\hbar\mu$.
 
 In order to look for the stationary solutions of 
 $$\left[\frac{1}{N(r)}\frac{\partial^2}{\partial t^2}-\frac{\partial}{\partial r}N(r)\frac{\partial}{\partial r}+\mathcal{U}_l(\mu, M; r)\right]\psi_{lm}=0,$$
 a further decomposition of the functions $\psi_{lm}(t,r)$ was done into oscillating modes of the form:
 $$\psi_{lm}(t, r)=e^{i\omega_{lm}t}u_{lm}(r),$$
 with $\omega_{lm}$ a real frequency and $u_{lm}(r)$ a complex function of r in the interval $(2M,\infty)$. 

Although stationary solutions where found for the scalar field, 
$$||u||^2:=\int^{\infty}_{2M}\left(N(r)\left|\frac{\partial u}{\partial r}\right|^2+\mathcal{U}_l(\mu, M; r)|u|^2\right)dr$$
they have been shown to be unphysical, in the sense that their energy density integrates to infinity in a compact region just outside the event horizon. However, there seem to exist long-lasting, quasi-stationary solutions of finite energy, which are found by evolving initial data that was constructed by slightly modifying a particular subset of the stationary solutions. The solutions found so far show as an overall behavior an exponential energy decay, caused by scalar field leaking into the black hole, that in some cases can be very slow.

The stationary solutions where obtained by solving a time-independent Schr\"odinger-like equation
$$\left[-\frac{\partial^2}{\partial r^{*2}}+V_{eff}(r^*)\right]u(r^*)=\omega^2u(r^*),\mbox{  }-\infty<r^*<\infty,$$
with an effective potential
$$V_{eff}(r^*):=N(r)\mathcal{U}_{l}(\mu, M; r),\mbox{  }r=r(r^*).$$
and hence they where characterized only by the properties of the potential. This fact is strictly true for the stationary solutions, but interestingly it was found that the quasi-stationary solutions, for which the Schr\"odinger-like equation no longer holds, can also be characterized by the properties of that same effective potential. Then, the cases of interest where those in which the effective potential contained a local minimum given by the condition
$$M\mu^2r^3-l(l+1)r^2+3M(l^2+l-1)r+8M^2=0.$$
The existence of this minimum then depended solely on the combination of the parameters $M_{\mu}$ and $l$. Although none of the possible forms of the effective potential proposed allowed for bound states, the existence of the well was enough to allow for resonant states, which where the ones that where useful in constructing initial data that could give rise to long-lasting quasi-stationary configurations of finite energy.

As mentioned by the authors, it could be objected that in order to obtain the mentioned quasi-stationary solutions very particular initial data should be constructed. However, it seems that the crucial factor is the existence of the potential well. Even when starting with modified stationary solutions that are not resonant, after an initial abrupt energy loss, the late time behavior observed resulted very similar to that of the resonant quasi-stationary solutions. These solutions seemed to evolve as a combination of the resonant modes. As mentioned before, the value of $\mu$ for scalar field dark matter models is expected to be given approximately by $\hbar\mu = 10^{-24}$eV in physical units, which gives rise to effective potentials with a local minimum for all values of $l$.

When evaluating the characteristic times of the solutions it was found that the longest lasting configurations could last for thousands of years. Although cosmologically this is very short time, for technical reasons, the authors where only able to study cases with relatively large values of $M_{\mu}$. Noting how fast the characteristic times seemed to increase with decreasing $\mu$, it resulted unreasonable to expect that configurations with $M_{\mu}\sim10^{-6}$ could last for cosmological time-scales.

Some aspects of this study are still open for some improvement. For example, a self-gravitating scalar field has not yet been considered. Second, much smaller values of the parameter $\mu$, and much larger scalar field distributions, could be needed to do a more realistic representation of dark matter halos. The main difficulty in dealing with such configurations result in handling the very different scales numerically. Third, besides studying possible quasi-stationary or long-lasting configurations with an already existing black hole, it would also be interesting to consider more dynamical scenarios such as the formation and/or growth of the black hole and the possibility of survival of the scalar field afterwards. The results presented by Barranco et al. in \cite{b37} seem to indicate that it is indeed possible for scalar field halos around super-massive black holes to survive for cosmological time-scales.

\section{Conclusions}

In this work we have revisited an alternative DM paradigm of the Universe known as scalar field dark matter or Bose-Einstein condensate 
dark matter model. In this model a fundamental scalar field plays the role of dark matter. 

We have reviewed a large number of recent quantitative and qualitative results aimed at explaining a variety of trends seen in the SFDM/BEC model. These trends include a brief description of Bose-Einstein condensates as dark matter, analysis and growth of its cosmological perturbations, its effect on the dynamics of galaxies, density profiles and the mass power spectrum, among others.

The key parameters of an ultralight BEC dark matter model are naturally shown to be the mass of the boson, which must be extremely 
small and, for the self-interacting scenario, the coupling strength of two-body repulsive interaction among the condensed particles. 
Dark matter is then suggested to arise from a single scalar field coupled to gravity undergoing a spontaneous symmetry break 
and hence rolls to a new minimum which gives a new vacuum expectation value. 
The breaking of symmetry can be done via a Ginzburg-Landau potential with quadratic and quartic terms. These give the mass and the 
interaction terms to the scalar field.

In the cosmological regime, it has been shown, that the SFDM/BEC model with an ultralight mass of $10^{-22}$eV mimics the behavior of the cosmological expansion rate predicted with the $\Lambda$CDM model. 

Another interesting cosmological behavior of the SF indicates that their scalar fluctuations can be appropriate for the purpose of 
structure formation,mainly because overdense regions of SFDM/BEC can produce the formation of galactic structure. Thus, the standard 
and the SFDM/BEC models can be contrasted in their predictions concerning the formation of the first galaxies. If in the future we 
observe more and more well formed and massive galaxies at high redshifts, this could be also a new indication in favor of the SFDM/BEC paradigm.

Much of the interest in this model has focussed on its ability to predict and agree with observations of the existence of dark matter 
and its capability to compete with the standard model of cosmology $\Lambda$CDM. Most of the themes we have described attempt to 
provide a wider view of what the model is and what is its current status. There are many other works in this field of research that are not less 
important, however, we trust that the interested reader will be able to deepen its knoledge with the references cited in this article. 

Clearly there are a large number of competing models to describe the dark matter of the Universe, and 
slight differences such as vortex formation or small shifts in cosmological perturbations may not be enough to decide which one is the 
best, however, detail observations give us a means to discard several of them and identify the ones that can stay. 

Finally, the SFDM/BEC model can have important implications in the nature of dark matter in the Universe. Additional work 
is needed if we are to fully understand this model, it would be desirable to have a unify framework that involves rotation curves, 
vortex formation, if present, and black hole effects. With all these intriguing results, we consider that the SFDM/BEC should be 
taken as a serious alternative to the dark matter problem in the Universe. The observational evidence seems to be in favor of 
some kind of cold dark matter, if we continue to observe even more galaxies at higher redshifts and if higher resolution observations 
of nearby galaxies exhibit a core density profile, this model can be a good alternative to $\Lambda$CDM. We expect that future 
observations in galaxy surveys can get us closer to the nature of dark matter.\\\\

{\bfseries\Large Acknowledgments}

This work was partially supported by CONACyt Mexico under grants CB-2009-01, no. 132400, no. 166212, and I0101/131/07 C-234/07 of the Instituto Avanzado de Cosmolog\'ia (IAC) collaboration (http://www.iac.edu.mx/).

\end{document}

%% file: econfmacros.tex
\def\beq{\begin{equation}}
\def\eeq#1{\label{#1}\end{equation}}
\def\eeqn{\end{equation}}

\def\beqa{\begin{eqnarray}}
\def\eeqa#1{\label{#1}\end{eqnarray}}
\def\eeqan{\end{eqnarray}}

\let\bar=\overbar

\def\Dslash{\not{\hbox{\kern-4pt $D$}}}
\def\dslash{\not{\hbox{\kern-2pt $\del$}}}

\def\msb{{\bar{\ssstyle M \kern -1pt S}}}

